\documentclass[fleqn,10pt]{wlpeerj}
\usepackage{url}
\usepackage{hyperref}
\title{Unsupervised Log Anomaly Detection with Few Unique Tokens}

\author[1]{Antonin Sulc}
\author[2,3]{Annika Eichler}
\author[2]{Tim Wilksen}
\affil[1]{Lawrence Berkeley National Lab, 1 Cycloton Rd., Bekreley, U.S.A.}
\affil[2]{Deutsches Elektronen-Synchrotron (DESY), Notkestr. 85, 22607 Hamburg, Germany}
\affil[3]{Technische Universitaet Hamburg (TUHH), Am Schwarzenberg-Campus 1, 21073, Hamburg, Germany}
\corrauthor[1]{Antonin Sulc}{asulc@lbl.gov}

\begin{abstract}
This article introduces a novel method for detecting anomalies within log data from control system nodes at the European XFEL accelerator. Effective anomaly detection is crucial for providing operators with a clear understanding of each node's availability, status, and potential problems, thereby ensuring smooth accelerator operation. Traditional and learning-based anomaly detection methods face significant limitations due to the sequential nature of these logs and the lack of a rich, node-specific text corpus. To address this, we propose an approach utilizing word embeddings to represent log entries and a Hidden Markov Model (HMM) to model the typical sequential patterns of these embeddings for individual nodes. Anomalies are identified by scoring individual log entries based on a probability ratio: this ratio compares the likelihood of the log sequence including the new entry against its likelihood without it, effectively measuring how well the new entry fits the established pattern. High scores indicate potential anomalies that deviate from the node's routine behavior. This method functions as a warning system, alerting operators to irregular log events that may signify underlying issues, thereby facilitating proactive intervention.
\end{abstract}

\begin{document}

\flushbottom
\maketitle
\thispagestyle{empty}

\section{Introduction}
The stability and reliability of the European XFEL (EuXFEL) facility are paramount for its successful operation. 
A network of hundreds of watchdog nodes continuously monitors the health of essential components, ensuring the proper functionality of crucial accelerator elements.
Given the volume and often redundant nature of logs generated by these nodes, manual monitoring of such specific streams of text can be challenging for humans and current rather large anomaly detection models. These logs, however, are invaluable, containing critical information about system status, including error messages and subtle anomalies that can indicate important insights about potential issues. Automating the analysis of these logs is therefore crucial for maintaining optimal node performance and overall facility uptime. This paper proposes a method leveraging language embedding and anomaly detection techniques to achieve this, enabling the early identification and mitigation of issues.
Detecting anomalies in EuXFEL log data presents unique challenges. The logs are often sparse, characterized by diverse but non-verbose entries. Compounding this, the system comprises hundreds of nodes, each potentially exhibiting distinct and minimalistic language patterns. 
These characteristics limit the effectiveness of many traditional or generic learning-based anomaly detection approaches that rely on rich, consistent textual corpora.
Our proposed method directly addresses these complexities. By modeling individual node log sequences using word embeddings and HMMs, we provide an automated means to comprehend EuXFEL system conditions. This enables the early detection and proactive resolution of subtle issues that might otherwise escalate or only become apparent after a significant node failure, thereby preventing undesired downtime and enhancing operational reliability.
This paper is structured as follows: Section~\ref{sect2} summarizes related work in log anomaly detection. Section~\ref{sect3} details the four main steps of our proposed approach, including key justifications and illustrative examples. Section~\ref{sect4} presents an evaluation of the method using EuXFEL log data. Finally, Section~\ref{sect5} concludes and outlines potential avenues for future work.
\subsection{Characterization of Anomalies}
Effective log analysis hinges on a precise definition of what constitutes an anomaly. For EuXFEL logs, we identify three primary categories of anomalies, each reflecting distinct deviations from expected system behavior:
\begin{itemize}
\item Sequential Anomalies: Log entries appearing out of their typical order or at unexpected points within a sequence. For example, a routine maintenance message logged during a critical operational phase.
\item Content Anomalies: Novel or rare log messages with content that significantly deviates from the established vocabulary of normal operations, such as unprecedented error messages or new system state indicators.
\item Contextual Anomalies: Standard log entries occurring in an unusual operational context or at an atypical frequency. An example is a normal status message appearing with abnormal regularity or during an incompatible system state.
\end{itemize}

\section{Related Work}
\label{sect2}
A common initial approach to log anomaly detection involves manually defined rule-based systems. For instance, Cinque et al. \cite{cinque2012event} and Yen et al. \cite{yen2013beehive} developed methods that scan logs for predefined patterns indicative of anomalies. However, these approaches rely heavily on expert knowledge to construct and maintain effective rules, a labor-intensive process that struggles with evolving systems. This limitation has spurred the development of more automated techniques leveraging machine learning.

With the rise of machine learning (ML), deep learning-based approaches have shown potential for thorough log analysis, particularly when a large log corpus is available, often accompanied by extensive, manually created labels. Long Short-Term Memory (LSTM) recurrent neural networks~\cite{du2017deeplog,zhang2016automated, zhang2019robust} have been popular for log anomaly detection due to their ability to handle sequential data. More recently, Transformer architectures~\cite{vaswani2017attention}, such as BERT~\cite{devlin2018bert}, have also been applied for this task \cite{guo2021logbert}. While powerful, the reliance of these deep learning models on vast training datasets and millions of parameters can limit their applicability in resource-constrained scenarios like ours at EuXFEL, where such large, labeled datasets are not readily available. For a broader survey of ML in log analysis, refer to~\cite{landauer2023deep}.

Addressing the need for automated feature representation, Bertero et al.~\cite{bertero2017experience} proposed treating logs as natural text, leveraging Word2Vec word vector representations~\cite{mikolov2013distributed,mikolov2013efficient} for automated embedding. This technique maps words to a vector space, allowing the use of off-the-shelf classifiers for anomaly detection. While this automates embedding, a major drawback is that their approach still relies on manual labeling to train the classifier, which can be prohibitively expensive. Furthermore, they treat each log entry independently, thereby ignoring the crucial sequential relationships between consecutive log messages.

To mitigate the need for labeled data, other works, such as \cite{lou2010mining} and \cite{xu2009online}, have explored unsupervised learning. These methods typically apply text mining techniques to logs and employ clustering algorithms to identify anomalies without manual labels. However, a common limitation is that they, too, often consider log entries in isolation rather than fully leveraging the contextual information embedded in log sequences.

In this work, we propose an alternative unsupervised approach designed to overcome these limitations, particularly the challenges posed by the limited diversity and volume of log entries at EuXFEL, which renders standard supervised ML models ineffective. Our method is engineered to adapt to novel log messages while fundamentally capturing the sequential nature of log data, without requiring labels or extensive user intervention.

Inspired by the embedding techniques in \cite{bertero2017experience}, we employ word vectors to represent log entries in a high-dimensional space, which helps mitigate data scarcity. However, crucially, instead of relying on supervised classifiers, we model the logs from each source as a sequence of events. Our key insight is to focus on modeling the patterns of co-occurring log sequences using an unsupervised technique based on Hidden Markov Models (HMMs) operating on these embedded sequences. By learning these sequential regularities, anomalies are detected as deviations from established contextual patterns, rather than solely from individual log content. This probabilistic HMM-based approach requires estimating only a minimal number of parameters, enabling robust anomaly detection even with limited and unlabeled training data.

\section{Method}
\label{sect3}
This section details our proposed method for scoring individual log entries to detect anomalies. The process consists of four primary stages:
\begin{itemize}
\item \textbf{Log Pre-processing}: Initially, raw log entries undergo a pre-processing stage designed to standardize their format and reduce noise. This involves consolidating redundant patterns and minimizing the impact of unique token sparsity, transforming the raw text into a consistent set of tokenized entries.
\item \textbf{Log Entry Embedding with Word2Vec}: Each pre-processed log entry is then converted into a dense numerical vector using Word2Vec. Specifically, we represent an entire log entry by calculating the mean of the Word2Vec vectors corresponding to its constituent tokens. This yields a fixed-length vector for each message, capturing its semantic content in a continuous vector space.
\item \textbf{Sequential Pattern Learning via HMM Optimization}: Subsequently, a Hidden Markov Model (HMM) is trained using sequences of these log entry embeddings derived from historical, observed logs. During this optimization phase with forward backward algorithm, the HMM learns a probability distribution over typical sequences of log entries, effectively modeling the normal temporal behavior and transitions between different types of log messages from a given source.
\item \textbf{Anomaly Scoring}: Finally, new, incoming log entries are scored based on their conformity to the trained HMM. As outlined in our abstract, we compute an anomaly score for an individual log entry as a probability ratio: this ratio compares the likelihood of the entire log sequence including the new entry against the likelihood of the sequence excluding it. A ratio indicating that the new entry significantly decreases the sequence's probability signifies a deviation from the learned sequential patterns, thereby identifying the entry as a potential anomaly.
\end{itemize}

\begin{figure}[ht!]
    \centering
    \includegraphics[width=1.0\linewidth]{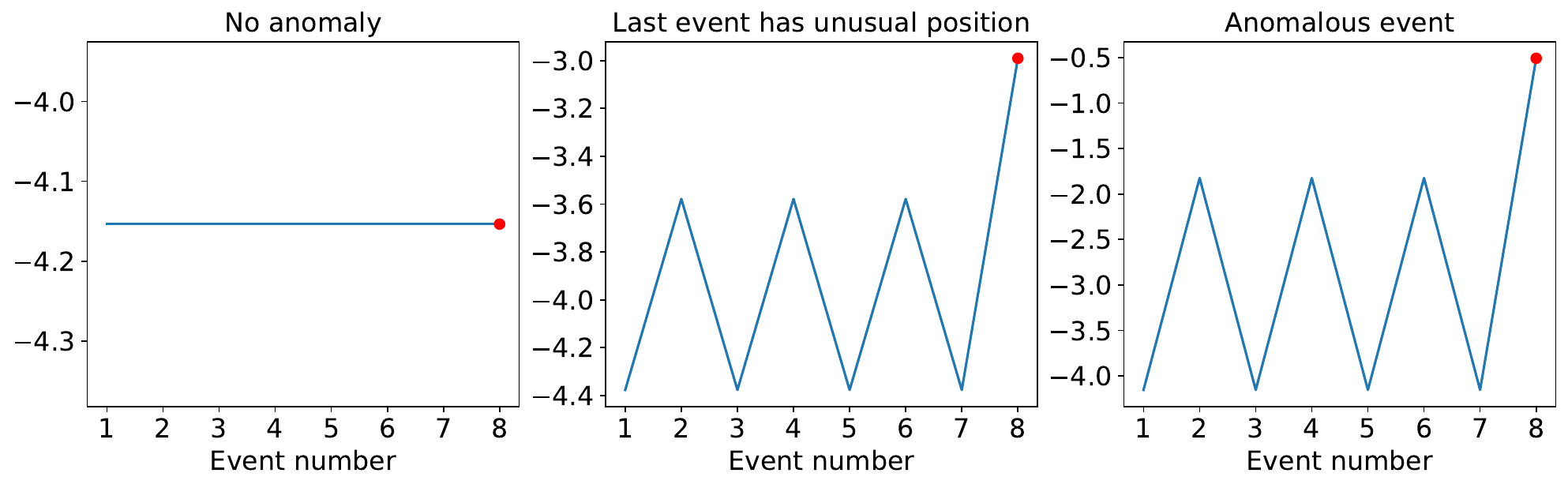}
    \caption{HMM Anomaly Detection Performance on a Synthetic Log Sequence: This figure demonstrates the proposed HMM's ability to detect anomalies in a synthetic log sequence. The sequence involves normal events ($o_1$,$o_2$) and a predefined anomalous event ($o_a$). For each scenario, the HMM parameters were estimated using the sequence except the evaluated entry (to ensure the model scores novel or altered final states without prior knowledge of them, minimizing overfitting for this test). The resulting anomaly score ($s$) for the final entry is observed: Left Panel (Normal Pattern): A sequence ending with the established, repetitive $o1$, $o2$ pattern. Value of $s$ remains consistently low, correctly identifying the sequence as normal. Center Panel (Disrupted Pattern): The sequence concludes with a swapped order ($o_2$, $o_1$), disrupting the expected $o_1$, $o_2$ pattern. This unexpected sequencing results in a noticeable increase in the $s$. Right Panel (Anomalous Event): The sequence terminates with the insertion of the novel, anomalous event $o_a$. This causes a substantial spike in the $s$, clearly highlighting the improbable observation. These scenarios illustrate that the HMM effectively detects anomalies stemming from both unexpected sequences of otherwise normal events (center) and the introduction of entirely novel or improbable log messages (right). Increases in the $s$ directly correlate with the degree of deviation from learned sequential patterns.}
    \label{fig1}
\end{figure}

\subsection{Preprocessing and Tokenization}
This section details the preprocessing pipeline applied to raw log text to prepare it for subsequent analysis. These steps are crucial for standardizing the log entries and reducing data sparsity. The process is as follows:
\begin{itemize}
\item Log Entry Segmentation: Individual log entries are first segmented from the continuous raw log messages. This separation is achieved by identifying standard delimiters such as timestamps at the beginning of an entry and newline characters marking its end.
\item Tokenization: Each segmented log entry is then tokenized into its constituent words or symbols using the NLTK punkt tokenizer \cite{bird2009natural}.
\item Token Normalization and Abstraction: Following tokenization, a series of transformations are applied to each token to create a uniform and abstracted representation:
\begin{enumerate}
\item Lowercasing: The entire log entry, and consequently all its tokens, are converted to lowercase to ensure case-insensitivity in subsequent processing.
\item Special Character Filtering: Special characters are removed. 
\item Entity Masking: Tokens potentially representing server or device names are replaced with generic placeholders. This includes tokens starting with facility-specific nomenclature (e.g., xfel, desy – after lowercasing) or those ending with common suffixes like svr or server or specific paths. 
\item Numeric Abstraction: Numeric tokens are categorized and replaced with placeholders to generalize numerical values: non-zero numbers become \texttt{\$nz} and zeros are replaced with \texttt{\$zero}.
\item Stopword Removal: Common English stopwords (e.g., "is", "a", "the"), as defined by the NLTK library \cite{bird2009natural}, are removed to reduce noise and focus on more content-rich tokens.
Embedding
\end{enumerate}
\end{itemize}

\subsection{Embedding}
To represent log entries as dense vectors, our approach utilizes Word2Vec~\cite{mikolov2013distributed,mikolov2013efficient}. The fundamental principle of Word2Vec is that words appearing in similar textual contexts are likely to possess similar semantic meanings. It operationalizes this by training a shallow neural network to learn word embeddings that capture these contextual semantic relationships. For this purpose, we employ the Continuous Bag-of-Words (CBOW) architecture, as introduced in~\cite{mikolov2013efficient}, which learns to predict a target word from its surrounding context words (an alternative to the skip-gram model, which predicts the context from a target word).

The resulting Word2Vec mapping produces vectors where proximity in the embedding space corresponds to semantic similarity. 
When applied to log analysis, Word2Vec can effectively learn and represent the relationships between terms that frequently co-occur, thereby capturing the intrinsic context within log messages. A significant capability of these learned embeddings is their support for meaningful arithmetic operations. For example, performing vector addition such as 'disk' and 'space' can result in a vector near those representing related concepts like ‘available’ or ‘lack’. Similarly, as illustrated in Fig. 2, combining the embeddings for linux and mac can yield a vector close to those of other operating system terms like windows and os.

This additive property is particularly valuable for generating a single vector representation for multi-word log entries. In our method, we achieve this by calculating the mean of the Word2Vec embeddings for all tokens within a given log entry. While more complex pooling strategies exist~\cite{gao2021simcse,reimers2019sentence}, this mean pooling approach proved both effective and sufficient for our requirements, providing a concise yet semantically rich representation for each log message.

\begin{figure}[ht!]
    \centering
    \includegraphics[width=1.0\linewidth]{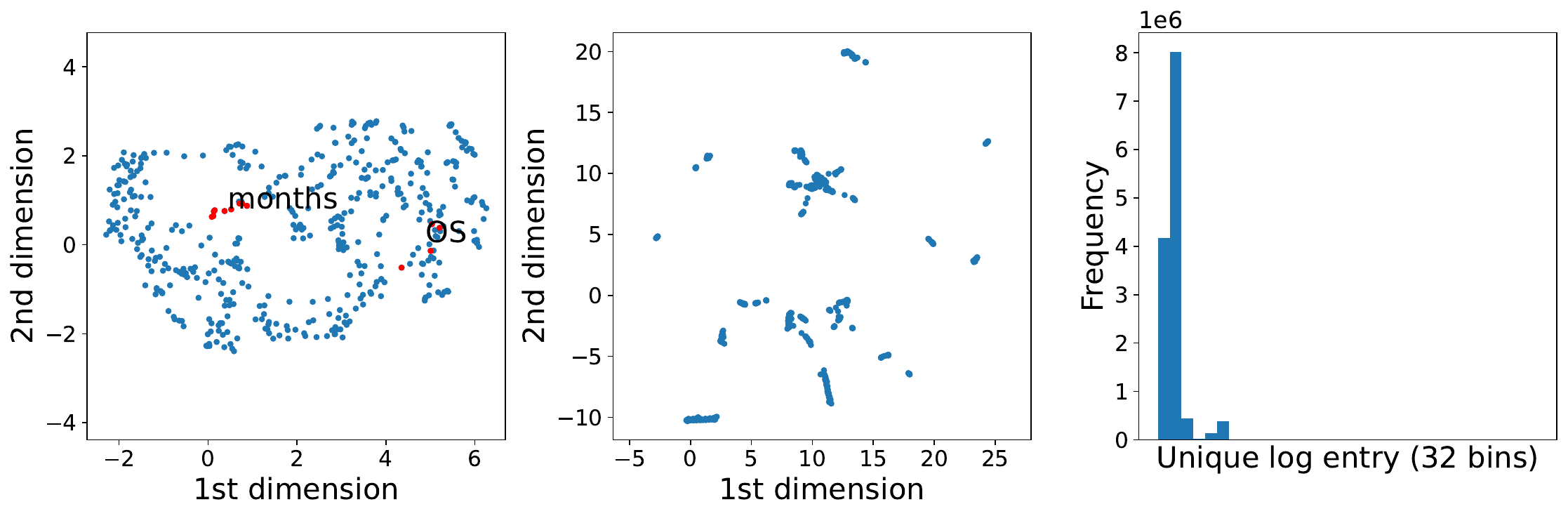}
    \caption{(Left) A 2D UMAP~\cite{mcinnes2018umap} projection of the 16-dimensional Word2Vec embeddings for all words in the vocabulary. Semantically similar words form distinct clusters, highlighted by red ellipses (or markers/color, specify how they are distinguished). Examples include clusters for three-letter month names, and operating system terms (e.g., 'windows', 'linux', 'mac', 'os'). This demonstrates that the Word2Vec model successfully captured latent semantic relationships even under these very limited conditions. (Center) A 2D UMAP projection of unique log entries, where each entry is represented by the sum of its 16-dimensional token embeddings. The non-uniform distribution indicates that log messages predominantly form a few distinct clusters, rather than being widely dispersed. (Right) Frequency distribution of unique, pre-processed log entries. The distribution exhibits a strong positive skew, indicating that a very limited set of messages accounts for the vast majority of log occurrences.}
    \label{fig2}
\end{figure}

\subsection{Anomaly detection with HMMs}
Adopting notation similar to~\cite{daniel2007speech}, let $Q = \{q_1, ..., q_N\}$ represent a set of $N$ hidden states, and $O = \left(o_1, ..., o_T\right)$ denote a sequence of $T$ observations. HMMs are characterized by two fundamental assumptions:
\begin{enumerate}
\item First-Order Markov Assumption: The current hidden state $q_t\in Q$ (at time $t$) is dependent only on the immediately preceding state $q_{t-1}$. Mathematically, this is expressed as 
$$P\left(q_t | q_1, \dots, q_{t-1}\right) = P\left(q_t | q_{t-1}\right).$$
\item Output Probability Assumption: The probability of observing $o_t$ at time $t$ depends solely on the hidden state $q_t$ at that same time, and not on any other hidden states or observations. This is formally
$$P\left(o_t | q_1, \dots, q_T, o_1, ..., o_{t-1}, o_{t+1} ..., o_T\right) = P\left(o_t | q_t\right).$$
\end{enumerate}

In our method, each observation $o_t$ in the sequence $O$ corresponds to the vector representation of an individual log entry. These vectors are the result of the preprocessing, tokenization, and $H$-dimensional embedding stages previously detailed. The hidden states $q_t\in Q$ are interpreted as the unobserved, underlying operational states of the system that generates these log entries.
Given a sequence of observed log vectors $o_1, ..., o_{t-1}$, our objective is to evaluate a new, incoming log vector $o_t$. Specifically, we aim to determine how probable its occurrence is in the context of the preceding sequence. This often involves comparing the probability of the complete sequence including $o_t$, i.e., $P\left(o_1,...,o_{t-1},o_t\right)$, with the probability of the sequence without it, $P\left(o_1,...o_{t-1}\right)$, to assess the marginal contribution or surprise of $o_t$
\begin{equation}
    s_i = \log \frac{P_\theta\left(o_1,\dots o_{t-1}\right)}{P_\theta \left(o_1,\dots o_t\right)} = \log P_\theta\left(o_1,\dots o_{t-1}\right) - \log P_\theta\left(o_1,\dots o_t\right)
\end{equation}

The anomaly score si quantifies the anomaly level of a new log entry $o_i$, based on the HMM parameters $\theta$. 
These parameters can be estimated from the sequence of log entries preceding $o_i$, i.e., $o_1,...o_{i-1}$ or from a relevant subsequence (e.g., using a sliding window). 
Parameter estimation methods are detailed in the following section.

A key requirement in log anomaly detection is the ability to handle novel entries~\cite{zhang2019robust}. 
While our method's generalization to logs containing entirely new, out-of-vocabulary terms is constrained by the initial embedding vocabulary, its emphasis on sequence modeling over individual word semantics is a core strength. This focus allows it to detect anomalies primarily through contextual irregularities and unexpected variations in log sequences, rather than relying solely on the content of individual messages. Indeed, we have observed that anomalies in our domain often manifest as unusual sequences of known events rather than just the appearance of specific unknown terms.
By basing the anomaly score si on the likelihood of an observed sequence under the learned HMM (where significant deviations from expected sequences, implying lower likelihoods, result in higher anomaly scores), rather than on predefined keyword rules, our approach can effectively evaluate previously unseen sequences of known log messages for contextual anomalies. This unsupervised, sequential methodology, which assesses log entries based on their contextual fit within learned patterns rather than relying on predefined labels, distinguishes our technique from supervised classification approaches such as those in~\cite{bertero2017experience,guo2021logbert,du2017deeplog,zhang2016automated,zhang2019robust}. However, it is important to acknowledge that false alarms may still occur if natural system fluctuations also produce sequences that deviate significantly from the learned patterns, an aspect we illustrate further in~\ref{fig7}

\subsection{Computational Complexity}
The computational complexity of our HMM implementation is analyzed for its two main phases: training and inference. For training, the Baum-Welch algorithm, used for estimating the model parameters $\theta$, has a time complexity of $O\left(TN^2\right)$ per iteration, where $T$ is the length of the observation sequence and $N$ is the number of hidden states. In our practical application, $N$ is typically small (between 2-8 states), which helps to keep the impact of the $N^2$ factor manageable.

For inference, such as computing the log-probability of an observation sequence using the Forward algorithm, the time complexity is also $O\left(TN^2\right)$.

A key advantage of the proposed HMM-based method is its memory efficiency. The model parameters require $O\left(N^2\right)$ storage space for the transition probabilities and $O\left(NH\right)$ for the emission probability parameters (assuming Gaussian emissions with diagonal covariance per state), where $H$ is the dimension of our log entry embeddings. 
The initial state probabilities require an additional $O(N)$ space. This compact parameterization is particularly beneficial for our application.

\subsection{Parameter Estimation}
For the real-world results presented (Figures~\ref{fig4},~\ref{fig5},~\ref{fig6}, and~\ref{fig7}), HMM parameters were consistently estimated using all available log messages that occurred prior to a designated test period. The log messages within this subsequent period were then exclusively used as the test set, ensuring a strict separation between training and evaluation data.

This strategy, where the training data comprises all historical log messages preceding the test window, proved computationally manageable on our hardware. Such efficiency is supported by the characteristics of the Baum-Welch algorithm, which is employed for HMM training; its time complexity scales linearly with the length of the input sequence ($T$), given a fixed number of hidden states ($N$).

\subsection{Handling Long and Noisy Sequences}
For stations generating extensive log data, such as those with tens of thousands of messages, the computational cost of HMM training can become substantial. To address this scalability challenge, we explored two distinct strategies to make training more manageable:
\begin{enumerate}
\item Sliding Window Training: This approach involves training the HMM on a moving subset (window) of the complete log sequence (conceptually similar to analyzing sequence segments as depicted in Fig.~\ref{fig1}). The relatively small number of parameters in our HMMs, owing to a limited number of hidden states ($N$), allows for effective model estimation even from these shorter subsequences.
\item Incremental Training with Parameter Initialization: Instead of retraining the HMM from default initializations for each new data segment or window, this strategy uses the parameters learned from a previous training phase as a warm start. This heuristic technique can accelerate convergence and help maintain parameter stability by preventing drastic deviations from established estimates. This can eventually be combined with (1).
\end{enumerate}

The sliding window method allows the model to dynamically focus on recent temporal contexts, while incremental training leverages prior learning to guide current parameter estimation. Both strategies offer pathways to maintain reasonable training times for very long sequences, though they introduce their own trade-offs (e.g., potential loss of longer-range dependencies with narrow windows) that we plan to evaluate more thoroughly in future work. While full sequence training was adequate for our initial datasets, these adaptive training techniques represent promising directions for handling larger-scale log data.
It is also crucial to consider an inherent characteristic of HMMs: performance on very long sequences can be affected by the exponential growth in the number of possible hidden state paths, which is $NT$ (where $N$ is the number of states and $T$ is the sequence length). This proliferation of paths can lead to diminishing certainty in state estimations and path probabilities over extended durations, as minor inaccuracies in model parameters may compound. However, this challenge is substantially mitigated by employing sliding window techniques over appropriately sized windows. This approach is particularly effective when the state space ($N$) is small, as show in Section~\ref{sect5}.

\begin{figure}[ht!]
    \centering
    \includegraphics[width=1.0\linewidth]{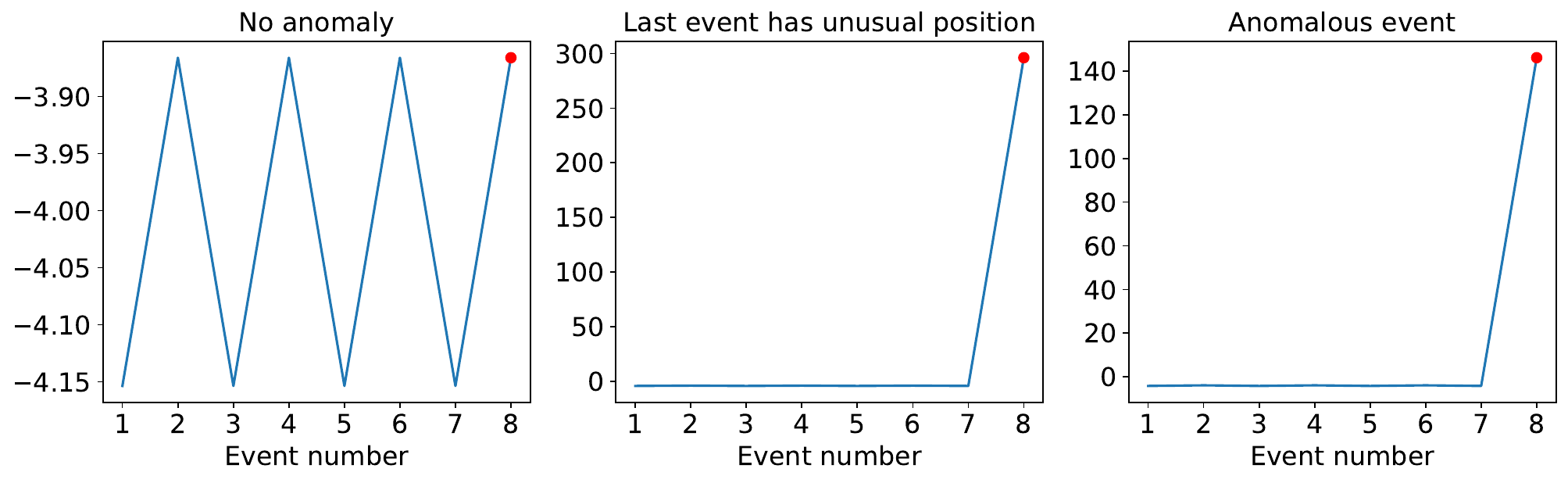}
    \caption{This figure demonstrates the HMM's ability to detect anomalies even when events of the anomalous type were included (infrequently as they are anomalous) during the parameter estimation (training) phase. The experiment utilizes synthetic log sequences identical in structure to those presented in Figure~\ref{fig1} (comprising normal events $o_1$, $o_2$, and an anomalous event $o_a$). However, unlike the scenario in Figure~\ref{fig1} where anomaly was excluded from parameter estimation, here the HMM is trained on the entire sequence, including these infrequent anomalous $o_a$ events. The results show that the model still effectively identifies anomalies—whether they manifest as unlikely novel messages (like $o_a$) or as unexpected sequences of normal messages. This occurs even though data containing such anomalies contributed to the HMM's parameter estimation, highlighting the model's capacity to detect deviations from the dominant learned patterns.}
    \label{fig3}
\end{figure}

In contrast to the idealized scenario depicted in Fig.~\ref{fig1} (where anomalies were deliberately excluded from the HMM training data), Fig.~\ref{fig3} illustrates the model's performance when its parameters ($\theta$) are estimated from a log sequence that includes anomalous entries. Remarkably, even with such 'contaminated' training data, the HMM effectively identifies events that significantly disrupt the predominantly learned sequential patterns. This capability is evidenced by pronounced spikes in the anomaly scores when these pattern-breaking events—whether they are new instances of anomaly types seen rarely during training or other forms of deviation—occur during the evaluation phase.

This highlights a crucial strength of HMM approach for anomaly detection: the ability to detect anomalies even if the training dataset is not perfectly 'clean.' Because the method emphasizes deviations from learned probabilistic sequences rather than matching specific content signatures, the presence of some irregular entries within the training set does not necessarily prevent the detection of subsequent events that are contextually improbable relative to the dominant learned patterns. This resilience is particularly valuable for post-mortem analysis scenarios or in operational environments where curated, anomaly-free training datasets are often unavailable. While the HMM might assign some (albeit low) probability to rare anomalous transitions if they were part of training, new occurrences that significantly disrupt common, high-probability sequences can still be effectively flagged.

\subsection{Example - Sequence Anomaly Detection}
To demonstrate how our HMM approach detects diverse types of anomalies, consider a conceptual HMM with two hidden states, $q_1$ and $q_2$. We illustrate its capability through two primary scenarios involving different observable outputs:
\begin{itemize}
\item Novel/Rare Event Anomaly: The observable vocabulary includes two common outputs, $o_1$ and $o_2$, along with a distinct anomalous event, $o_a$, which is either entirely novel or very infrequent in normal operational data.
\item Sequential Anomaly: The observable vocabulary consists only of standard outputs, $o_1$ and $o_2$. However, one of these common outputs appears in an unexpected sequential position, violating the learned transition patterns.
\end{itemize}

In the first scenario, the appearance of the anomalous output $o_a$ is highly improbable under the HMM trained on normal data (i.e., it would have a very low emission probability from any likely hidden state, or involve low-probability state transitions to emit it). This low likelihood of occurrence results in a significantly increased anomaly score, thereby flagging $o_a$ as anomalous.

In the second scenario, even though the individual observed output ($o_1$ or $o_2$) is familiar, its occurrence in an unlikely position (e.g., following an improbable state transition according to the learned sequence dynamics) makes the overall sequence less probable. This deviation from the expected sequence also leads to an increased anomaly score.

To illustrate these principles, we created a minimal 8-event synthetic example (detailed in Figures~\ref{fig1} and~\ref{fig3}) where disrupting an established pattern clearly impacts the anomaly scores. Critically, for the primary demonstration in Figures.~\ref{fig1} and~\ref{fig3}, the HMM parameters were estimated excluding the final anomalous event being tested.  Furthermore, as shown in Fig.~\ref{fig3}, the HMM retains its ability to detect such disruptions even if data containing similar (infrequent) anomalous events was included during parameter estimation. These examples demonstrate that the model identifies anomalies by recognizing significant deviations from learned sequential patterns, highlighting its potential even with limited or somewhat 'contaminated' input data.

In essence, our HMM approach can detect anomalies arising either from the content of log messages (unlikely or novel events) or from their context (standard messages appearing in surprising positions that break expected sequence patterns). The HMM identifies observations or sequences that diverge significantly from the learned probabilistic distributions, assigning higher anomaly scores to these improbable occurrences. The examples in Fig.~\ref{fig1} and Fig.~\ref{fig3} not only confirm these diverse detection capabilities but also suggest that the foundational parameters of a simple HMM can be reasonably estimated from a limited number of events, and the model can show resilience even when the training input contains some irregularities.

\subsection{Analysis - Word Embedding}
This section provides a detailed analysis of the word embeddings generated from our log corpus. Our aim is to demonstrate Word2Vec's robustness in extracting meaningful semantic representations, even though log entries are not conventional natural language and our corpus presents specific challenges.

Processing these logs with Word2Vec is particularly insightful due to the corpus's distinct characteristics: it contains a very limited vocabulary (only 475 unique tokens after preprocessing and tokenization) and results in fewer than 1,000 unique log messages. The distribution illustrating the low count of unique messages is shown in Figure~\ref{fig2} (Right Panel). This pronounced lack of diversity in both the vocabulary and the unique message structures inherently constrains the feasibility of more complex, parameter-rich language modeling approaches.

Despite these limitations, Figure~\ref{fig2} (Left Panel) - the 2D UMAP projection of the individual word embeddings – demonstrates that Word2Vec successfully captures semantic relationships. As shown by highlighted examples in the projection, tokens with similar meanings or belonging to the same category (e.g., month names, specific user names, or operating system terms) are embedded in close proximity to one another. Furthermore, Figure~\ref{fig2} (Center Panel) visualizes the UMAP projection of log entry embeddings, which are derived by averaging the Word2Vec vectors of their constituent tokens. This panel further underscores the challenge posed by the limited diversity of log entries, revealing that the entry vectors themselves form only a few densely packed clusters, reflecting the repetitive nature of many log messages.

\subsection{Implementation}
The implementation was developed using Python version 3.9. For the generation of log message embeddings via Word2Vec, we utilized the Gensim library~\cite{rehurek2010gensim}. Sequence modeling with Hidden Markov Models was performed using the Hmmlearn package\cite{hmmlearn}.

\section{Experiments and Results}
To our knowledge, there is no standardized dataset currently available for this specific purpose. Due to privacy constraints, we are unable to release our full datasets, with the exception of a few minimal examples.
All examples included in this paper have been thoroughly vetted to ensure that no private or sensitive information is revealed. To showcase the efficacy of our proposed method, we have carefully chosen and presented a number of representative examples.
The evaluation of these examples was conducted by human operators.

For evaluation, we used a setup with Intel(R) Core(TM) i7-11850H @ 2.50GHz and 16GB. No acceleration was used.

\begin{figure}[ht!]
\begin{tabular}{l}
\begin{tabular}{ll}
1 & \texttt{remoteerrors errorcount \$nz}\\
2 & \texttt{remoteerrors errorcount \$nz}\\
3 & \texttt{remoteerrors errorcount \$nz}\\
4 & \texttt{remoteerrors errorcount \$nz}\\
5 & \texttt{remoteerrors errorcount \$nz toggled \$nz times \$nz min}\\
6 & \texttt{remoteerrors errorcount \$nz}\\
7 & \texttt{remoteerrors errorcount \$nz}\\
8 & \texttt{remoteerrors errorcount \$nz}\\
9 & \texttt{remoteerrors errorcount \$nz}\\
10 & \texttt{remoteerrors errorcount \$nz toggled \$nz times \$nz min}\\
11 & \texttt{remoteerrors errorcount \$nz}\\
12 & \texttt{rpccheck nullproc error}\\
13 & \texttt{rpccheck fails \$nz kill \$nz}\\
14 & \texttt{getpid pid not match process name
no process try start}\\
15 & \texttt{getpid pid not match process name
pid change \$nz \$nz}\\
16 & \texttt{rpccheck nullproc error}\\
17 & \texttt{rpccheck fails \$nz kill \$nz}\\
18 & \texttt{rpccheck fails \$nz kill \$nz}
\end{tabular}\\
\includegraphics[width=1.0\linewidth]{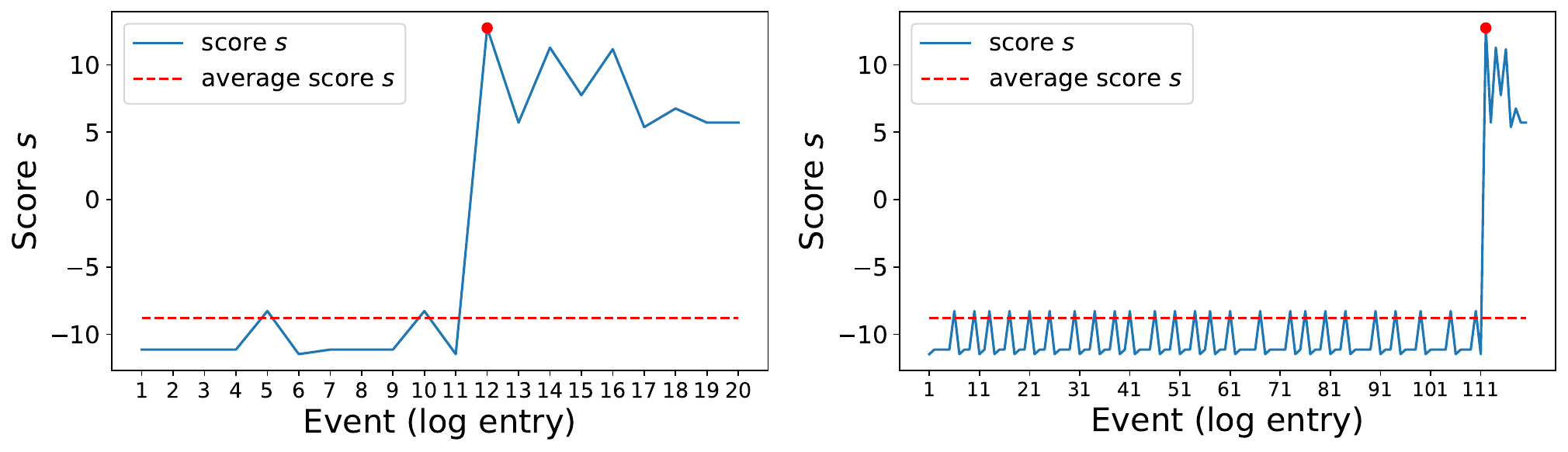}
\end{tabular}
\caption{An anonymized instance of an anomalous log (upper section), with anomalous events becoming evident starting from row 12. This is marked by the sudden appearance of the error message “rpccheck nullproc error”, coinciding with a notable increase in anomaly scores (depicted in the bottom left and right figures). The more detailed score plot (bottom left) provides a close-up view of the scores, revealing a rapid increase beginning at the 12th log entry, indicative of a significant error. The score plot in the bottom right showcases scores for the last 100 log entries. The dashed line shows an average score through the entire log}
\label{fig4}
\end{figure}

\begin{figure}[ht!]
\begin{tabular}{l}
\begin{tabular}{ll}
1 & \texttt{remoteerrors errorcount \$nz}\\
2 & \texttt{remoteerrors errorcount \$nz}\\
3 & \texttt{remoteerrors errorcount \$nz}\\
4 & \texttt{remoteerrors errorcount \$nz toggled \$nz times \$nz min}\\
5 & \texttt{rpccheck clnt create error}\\
6 & \texttt{remoteerrors errorcount \$nz}\\
7 & \texttt{rpccheck fails \$nz kill \$nz}\\
8 & \texttt{pid change \$nz \$nz}\\
9 & \texttt{rpccheck fails \$nz kill \$nz}\\
10 & \texttt{pid change \$nz \$nz}\\
11 & \texttt{rpccheck fails \$nz kill \$nz}\\
12 & \texttt{no process try start}\\
13 & \texttt{pid change \$nz \$nz}\\
14 & \texttt{remoteerrors errorcount \$nz}\\
15 & \texttt{no process try start toggled \$nz times \$nz min}\\
16 & \texttt{remoteerrors errorcount \$nz}\\
17 & \texttt{remoteerrors errorcount \$nz toggled \$nz times \$nz min}\\
18 & \texttt{remoteerrors errorcount \$nz}\\
19 & \texttt{remoteerrors errorcount \$nz}\\
20 & \texttt{remoteerrors errorcount \$nz toggled \$nz times \$nz min}\\
\end{tabular}\\
\includegraphics[width=1.0\linewidth]{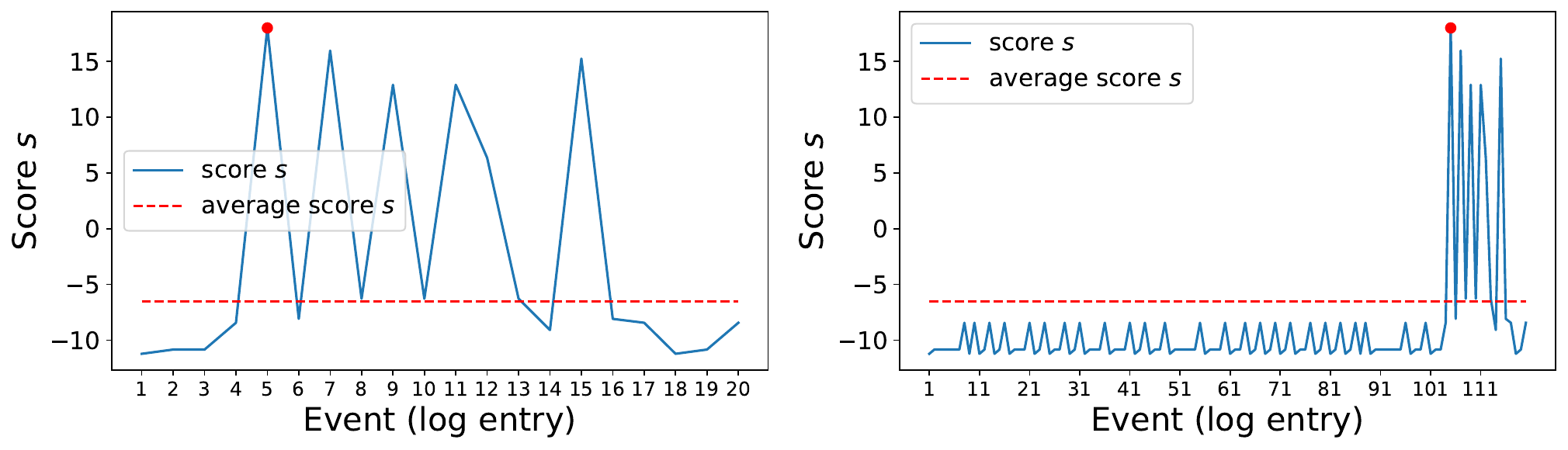}
\end{tabular}
\caption{An anonymized example of an anomalous log (top section), with the emergence of anomalous events starting at row 5. The anomalous series of events starts with the occurrence of the message \texttt{rpccheck clnt create error}. Consequently, an abrupt increase in anomaly scores is observed (visible in the bottom left). Notice the score plot in greater detail (bottom left) presents a sharp escalation in scores following the 5th log entry. Furthermore, despite log entry 6 bearing resemblance to previous entries, the score present in log entry 7 induces a slight increase in the score associated with log entry 5. The score chart (bottom right) provides an overview of scores for the preceding 100 log entries, prominently highlighting the location of the likely occurrence of the problem.
}
\label{fig5}
\end{figure}

\begin{figure}[ht!]
\begin{tabular}{l}
\resizebox{1.0\linewidth}{!}{
\begin{tabular}{ll}
1 & \texttt{doocs \$server \$path programmer ANONYMIZEDNAME ANONYMIZEDNAME}\\
2 & \texttt{doocs serverlib built on host mcsvdev ANONYMIZEDNAME on nov \$nz \$nz \$nz}\\
3 & \texttt{archiver runs advanced mode}\\
4 & \texttt{archiver uses histdir \$path}\\
5 & \texttt{started \$nz locations on \$server doocsarch ubuntux}\\
6 & \texttt{postinit phase completed}\\
7 & \texttt{getpid no process}\\
8 & \texttt{no process try start}\\
9 & \texttt{getpid no process}\\
10 & \texttt{getpid no process}\\
11 & \texttt{no process try start}\\
12 & \texttt{getpid no process}\\
13 & \texttt{no process try start}\\
14 & \texttt{no process try start}\\
15 & \texttt{pid change \$nz \$nz}\\
16 & \texttt{getpid pid not match process name}\\
17 & \texttt{pid change \$nz \$nz}\\
18 & \texttt{getpid pid not match process name}\\
19 & \texttt{pid change \$nz \$nz}\\
20 & \texttt{pid change \$nz \$nz}\\
21 & \texttt{pid not match process name toggled \$nz times \$nz min}\\
22 & \texttt{pid not match process name toggled \$nz times \$nz min}\\
23 & \texttt{signal term received}\\
24 & \texttt{terminating threads closing files}\\
25 & \texttt{writer thread terminated}\\
26 & \texttt{interrupt thread terminated}
\end{tabular}}\\
\includegraphics[width=1.0\linewidth]{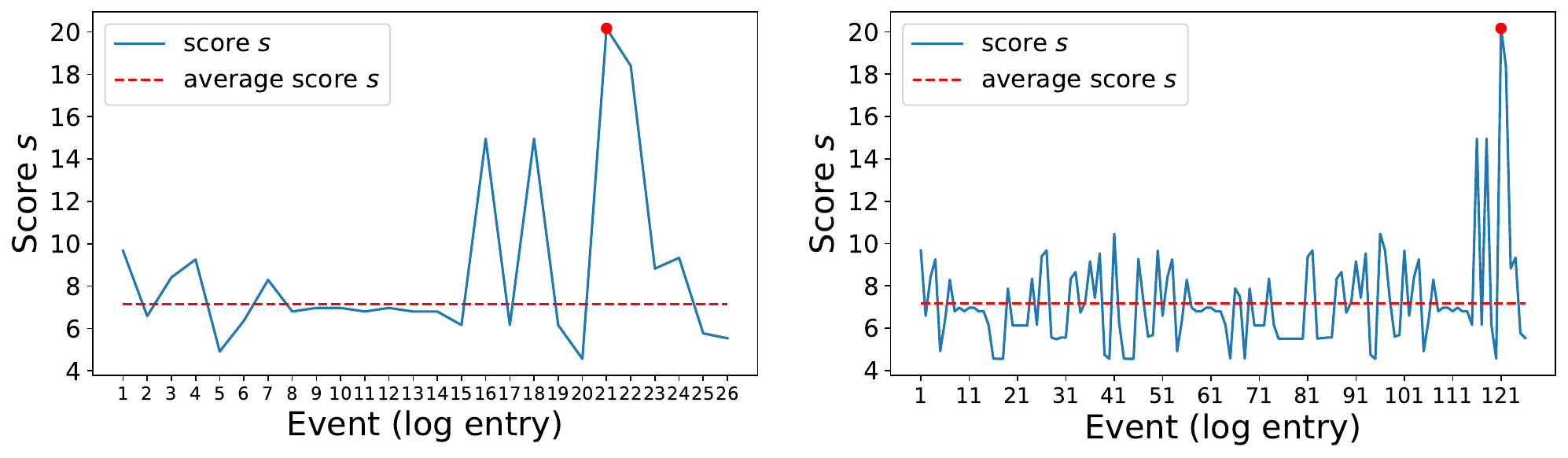}
\end{tabular}
\caption{In this anonymized log, an anomaly appears to occur at row 21. At this point, the error message \texttt{pid not match process name toggled \$nz times \$nz min} appears twice. This corresponds to a spike in the anomaly scores, as seen in the bottom left chart. However, similar spikes occur earlier at log entries 16 and 18 (\texttt{getpid pid not match process name}), suggesting other potential anomalies. When viewed in the context of the previous 100 log entries, the score spike at 16 is prominent, as illustrated in the bottom right chart. 
}
\label{fig6}
\end{figure}

\begin{figure}[ht!]
\begin{tabular}{l}
\begin{tabular}{ll}
1 & \texttt{config file create error}\\
2 & \texttt{config file create error}\\
3 & \texttt{config file create error}\\
4 & \texttt{remoteerrors errorcount \$nz}\\
5 & \texttt{config file create error}\\
6 & \texttt{config file create error}\\
7 & \texttt{config file create error}\\
8 & \texttt{config file create error}\\
9 & \texttt{rpccheck nullproc error}\\
10 & \texttt{no process try start}\\
11 & \texttt{pid change \$nz \$nz}\\
12 & \texttt{getpid no process}\\
13 & \texttt{no process try start}\\
14 & \texttt{pid change \$nz \$nz}\\
15 & \texttt{getpid no process}\\
16 & \texttt{no process try start}\\
17 & \texttt{pid change \$nz \$nz}\\
18 & \texttt{rpccheck nullproc error toggled \$nz times \$nz min}\\
19 & \texttt{rpccheck nullproc error}\\
20 & \texttt{rpccheck fails \$nz kill \$nz}
\end{tabular}\\
\includegraphics[width=1.0\linewidth]{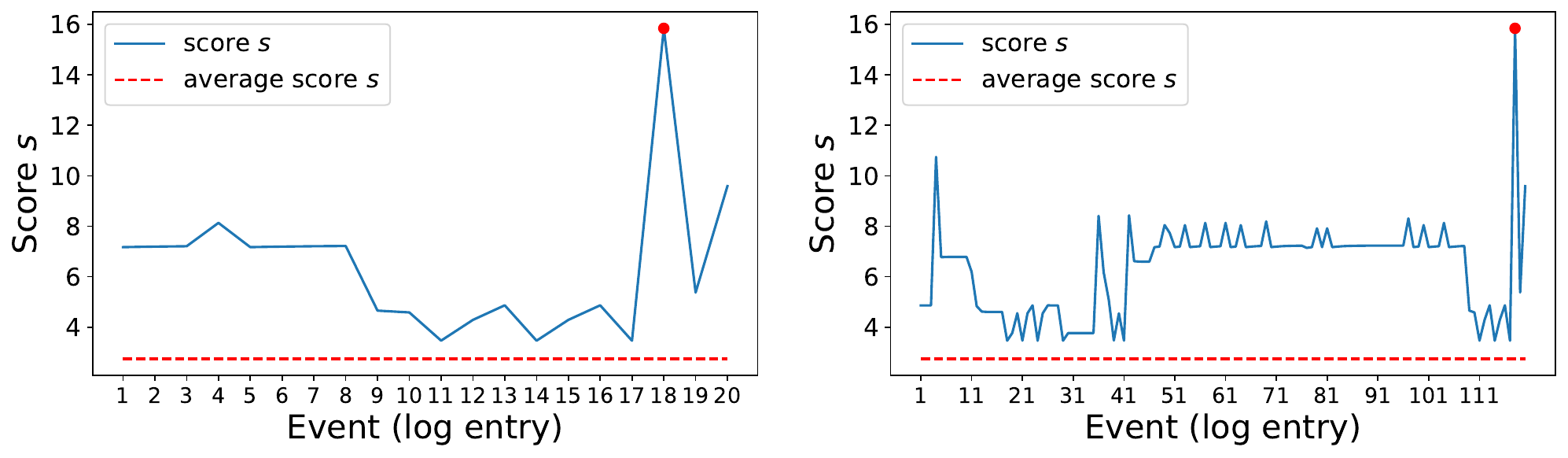}
\end{tabular}
\caption{In this anonymized log, an anomaly appears at row 18, where the error message \texttt{rpccheck nullproc error toggled \$nz times \$nz min} first appears. This corresponds to a spike in the anomaly scores, as shown in the bottom left chart. Interestingly, earlier error messages like “config file create error” have relatively high anomaly scores as well. However, since these repeat multiple times, their scores are slightly lower than the message 18. The message \texttt{pid change \$nz \$nz} also appears repeatedly in different contexts. Typically, this message does not indicate anything unexpected, which is why its anomaly scores tend to be much lower on average. The spike at row 18 stands out as the most prominent anomaly in this log example.	The bottom right chart shows the anomaly scores for the last 100 log messages from this node. It illustrates that this node tends to produce various (not necessarily) error messages fairly often, leading to generally high anomaly scores and many false positives with our method.}
\label{fig7}
\end{figure}

\label{sect4}
To demonstrate the performance of our anomaly detection method, this section presents an analysis of four representative instances. The input logs and their corresponding computed anomaly scores for these instances are illustrated in Figures~\ref{fig4},~\ref{fig5},~\ref{fig6}, and~\ref{fig7}.

Figure~\ref{fig4} displays a segment of an input log where the method performs as anticipated. The sudden appearance of the \texttt{rpccheck nullproc error} message at row 12 directly coincides with a sharp increase in the calculated anomaly scores. This increase is clearly visible in the accompanying charts (bottom left and right), and the distinct spike exemplifies the method's capability to detect unambiguous anomalous events.

Similarly,~Figure~\ref{fig5} provides another example of successful and expected anomaly detection. In this case, an error message occurring at row {5} triggers an abrupt escalation in the anomaly scores, as depicted in the bottom left chart. This observation further confirms the method’s proficiency in identifying potential system issues.

Figure~\ref{fig6} illustrates a more nuanced scenario involving greater operational complexity. Score spikes observed at rows 16, 18, and 21 indicate multiple events flagged as potential anomalies by the HMM. However, when contextualized by the preceding 50 entries (visualized in the bottom right chart), the spike at row 14 emerges as the most prominent. This case highlights that while the method can identify several potential irregularities, subsequent verification or contextual analysis may be necessary to pinpoint the most critical anomaly among them.

Finally, Figure~\ref{fig7} explores the method's performance under challenging conditions, revealing some inherent limitations. For this particular log, a consistently high baseline level of anomaly scores makes it more difficult to distinguish significant anomalies from the typical background 'noise.' Furthermore, frequent minor errors can contribute to a higher rate of potential false positives. Despite these difficulties, a discernible increase in the anomaly score at row 18 suggests that the method can still indicate likely anomalies, particularly when results are examined closely or relative changes above the baseline are considered.

In summary, the proposed method reliably flags issues when errors manifest as clear, substantial spikes in anomaly scores, as demonstrated in the examples in Figure~\ref{fig4} and Figure~\ref{fig5}. In more ambiguous cases presenting multiple potential signals, such as in Figure~\ref{fig6}, the outputs may necessitate further validation to identify the most significant events. For completeness, Figure~\ref{fig7} illustrates a scenario where performance is less definitive due to a noisy data environment; however, even under such conditions, salient anomalies can often be discerned by analyzing the patterns and relative changes in the anomaly scores.

\subsection{Infrasctucture}

\section{Future Work}
\label{sect5}
While this study establishes foundational anomaly detection capabilities using HMM on log data, several promising avenues exist for future enhancement and expansion.

One immediate direction is the exploration of more advanced or complementary anomaly detection techniques, such as those by~\cite{gornitz2015hidden}, which could potentially offer improved precision or recall. Furthermore, enhancing the verbosity of log data, if feasible, could generate richer datasets. This, in turn, might enable the effective deployment of more complex, parameter-rich algorithms, addressing a common challenge with sparse log information highlighted in related work.

A significant opportunity for advancing detection performance lies in the integration of heterogeneous data sources beyond textual log messages. Valuable diagnostic information is often contained within operational metrics such as CPU utilization, memory consumption, network traffic statistics, and disk I/O patterns. Effectively fusing these asynchronous, numerical time series with the sequential nature of log messages presents modeling challenges. Developing robust algorithms capable of jointly analyzing these diverse data streams to create a better view of system state provides an interesting opportunity for more insights. 

Finally, given the increasing prevalence of cybersecurity threats, adapting and extending this anomaly detection framework for security-specific use cases is considerable. By combining insights from log analysis with detailed knowledge of the specific infrastructure and potentially incorporating network traffic flow data, specialized models could be developed to identify patterns indicative of security incidents, intrusions, or vulnerabilities with special consideration. 

\section{Conclusion}
This paper presented a novel unsupervised approach for anomaly detection in log data. Our method utilizes Word2Vec embeddings to generate numerical representations of log entries and subsequently employs Hidden Markov Models (HMMs) to model their sequential behavior. Anomalies are identified by evaluating the statistical likelihood of new log messages within the context of previously observed sequences; significant deviations from these learned patterns manifest as prominent increases in the calculated anomaly score.

Experimental results on real-world log data from the EuXFEL facility demonstrate the method's efficacy in flagging potential system issues. Such issues are typically indicated by salient spikes in the anomaly scores, corresponding to operational errors or disruptions of established patterns. Key strengths of this approach include its entirely unsupervised nature, which obviates the need for labeled training data, and its fundamental reliance on modeling the inherent sequential dynamics of individual node logs through HMMs.

While these findings are promising, we acknowledge that challenges persist, particularly in enhancing robustness to highly noisy log data and further minimizing false positives—aspects observed in certain log instances. Future efforts will aim to address these limitations. Nevertheless, this HMM-based technique offers a valuable tool for proactive system monitoring in complex operational environments.

The code and materials for this study are openly available in a GitHub repository at \href{https://github.com/sulcantonin/LOG\_ICALEPCS23}{https://github.com/sulcantonin/LOG\_ICALEPCS23}.

\section{Acknowledgement}
The authors wish to acknowledge DESY (Hamburg, Germany), a member of the Helmholtz Association (HGF), for providing the vital resources and infrastructure that supported this study. We are also very grateful to our colleagues in the MCS and MSK groups, the EuXFEL team, and EuXFEL management for their valuable contributions to this research and their help in preparing this paper.

An Grammarly and Gemini AI-powered language tool was employed to enhance the manuscript's grammar and clarity, without being used for text generation.

\bibliography{references}

\end{document}